\documentclass[a4paper]{jpconf}
\usepackage{graphicx}
\usepackage{epstopdf}

\begin{document}
\title{The STAR tracking upgrade}

\author{F Simon, for the STAR Collaboration}

\address{Massachusetts Institute of Technology, 77 Massachusetts Ave., Cambridge, MA 02139, USA}

\ead{fsimon@mit.edu}

\begin{abstract}
The STAR experiment at the Relativistic Heavy Ion Collider RHIC studies the new
state of matter produced in relativistic heavy ion collisions and the spin structure of the nucleon in
collisions of polarized protons. In order to improve the capabilities for heavy flavor measurements
and the reconstruction of charged vector bosons an upgrade of the tracking system both in the
central and the forward region is pursued. The challenging environments of high track multiplicity in heavy ion collisions and of high luminosity in polarized proton collisions require the use of new technologies. The proposed inner tracking system, optimized for heavy flavor identification, is using active pixel sensors close to the collision point and silicon strip technology further outward. Charge sign determination for electrons and positrons from the decay of $W$ bosons will be provide by 6 large-area triple GEM disks currently under development. A prototype of the active pixel detectors has been tested in the STAR experiment, and an extensive beam test of triple GEM detectors using GEM foils produced by Tech-Etch of Plymouth, MA  has been done at Fermilab.
\end{abstract}

\section{Introduction}

The STAR experiment at RHIC studies the fundamental properties of the new state of strongly interacting matter produced in relativistic heavy ion collisions and investigates the spin structure of the proton in polarized $p+p$ collisions. A variety of results both in heavy ion collisions and polarized $p+p$ collisions have already been obtained. A key future step in these programs is the ability for direct reconstruction of particles containing charm and bottom quarks as well as flavor tagging of jets to allow precise measurements of the spectra, yields and flow of open charm and bottom \cite{Adams:2005dq}, and to determine spin dependent production asymmetries connected to the gluon polarization in the nucleon \cite{Bunce:2000uv}. The flavor dependence of the sea quark polarization will be determined by parity violating $W$ production and decay in longitudinally polarized $p+p$ collisions at $\sqrt{s}$ = 500 GeV \cite{Bunce:2000uv}. 

STAR \cite{Ackermann:2002ad} is one of the two large detector systems at RHIC. Its main tracking detector is a large-volume time projection chamber (TPC) covering the pseudorapidity range $\vert \eta \vert < 1.2$. Additional vertex resolution for the reconstruction of secondary decay vertices is provided by the silicon vertex tracker (SVT, $\vert \eta \vert < 1$), a three--layer silicon drift detector, and the one--layer silicon strip detector (SSD). Tracking in the forward region is provided by the forward TPCs (FTPCs, $2.5 < \vert \eta \vert < 4.0$). The barrel (BEMC) and endcap (EEMC) electromagnetic calorimeters cover $-1 < \eta < 1$ and $1 < \eta < 2$, respectively. Forward electromagnetic calorimetry is provided by the forward meson spectrometer (FMS, $2.5 <  \eta  < 4.0$). 

\begin{figure}
\centering
\includegraphics[width=0.95\textwidth]{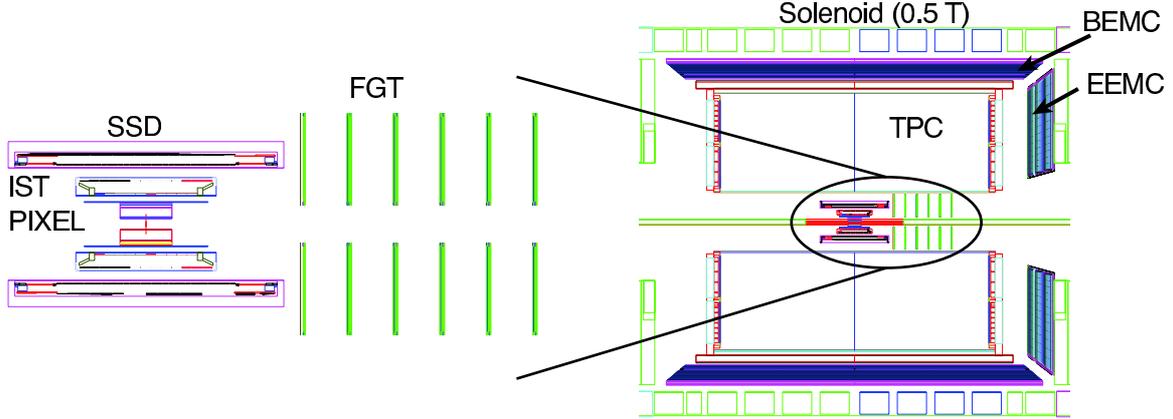}
\caption{Side view of the planned tracking upgrades in the STAR detector. The inner region is shown enlarged. The inner tracking system covering $\vert \eta \vert < 1$, referred to as the Heavy Flavor Tracker (HFT), consists of a two-layer silicon pixel detector (PIXEL), a two layer detector using back-to-back single sided silicon strip sensors (IST) and the existing double-sided silicon strip detector SSD. The Forward GEM Tracker (FGT) consists of six triple-GEM disks, covering $1 < \eta < 2$ over a wide range of collision vertices.}
\label{fig:Upgrades}
\end{figure}

The current tracking capabilities are insufficient to address the future measurements outlined above. The planned integrated tracker is designed to provide the necessary vertex resolution to uniquely identify open charm and bottom and to provide precision tracking in the forward region to determine the charge sign of electrons from $W^+$ and $W^-$ decays that are detected in the EEMC. Figure \ref{fig:Upgrades} shows an overview of the planned tracking upgrades for STAR. The two distinct areas of inner and forward tracking are driven by different physics motivations. The upgrade of the inner tracker, called Heavy Flavor Tracker (HFT) will provide the capability for topological reconstruction of open charm and open bottom, while the upgrade of the forward tracker, the Forward GEM Tracker FGT, will enable STAR to identify the charge sign of electrons and positrons produced in $W$ boson decays.

\section{The Heavy Flavor Tracker}

Open charm and open beauty reconstruction requires excellent vertexing to resolve the secondary decay vertex of these particles, since $c\tau\, \sim 120\ \mu\mbox{m}$ for $D^0$ and $c\tau\, \sim 460\ \mu\mbox{m}$ for $B^0$. The planned Heavy Flavor Tracker HFT  is designed to achieve this both in heavy ion collisions and in polarized $p+p$ collisions by an optimization for high multiplicity and high rate environments. A thin beryllium beam pipe with a radius of 2 cm will be used to give the detectors close access to the collision point. The HFT consists of three devices, all covering $\vert \eta \vert < 1.0$. The two innermost layers (PIXEL) are based on Active Pixel Sensors (APS) with 30 $\mu$m $\times$ 30 $\mu$m pixels using silicon thinned down to 50 $\mu$m, limiting the material of the detector to $\sim$ 0.3\% $X_0$ per layer, including cables and support. The inner sensor layer sits at a radius of 2.5 cm and a staggered outer layer sits at 6.5 cm and 7.5 cm radius. A fast intermediate tracker is needed to act as a pointing device from the TPC to the PIXEL to connect the precision points in the PIXEL to TPC tracks and to provide the time resolution necessary for high luminosity running. The gap to the existing double-sided silicon strip detector SSD at 23 cm will be bridged by the intermediate silicon tracker IST, consisting of two layers of conventional back-to-back silicon strip sensors at 12 cm and at 17 cm radius. The material budget for this fast device is estimated to be $\sim$ 1.5 \% $X_0$ per layer. The intermediate tracker will replace the SVT which does not have sufficient rate capability for future collider luminosities and is incompatible with ongoing upgrades of the STAR data acquisition system. The complete HFT will achieve an overall pointing precision of 40 $\mu$m to the primary event vertex. 

In the 2007 RHIC Au+Au run several small (4 mm $\times$ 4 mm) active pixel sensors were tested in the STAR experiment. The detectors were mounted at a distance of 5 cm from the beam, about 1.45 m from the interaction point, facing towards the interaction point. The detectors pick up the beam background and forward-going particles at radii comparable to that of the PIXEL detector layers. The data of this test is still being analyzed, but preliminary results indicate a stable performance of the the test modules, demonstrating the applicability of the technology in the environment of heavy ion collisions. 

\section{The Forward GEM Tracker}

At STAR, produced $W$s will be detected via their leptonic decays into an electron and a neutrino, $W^+ \rightarrow e^+ \nu_e$ and $W^- \rightarrow e^- \bar{\nu}_e$. The energy of the outgoing lepton will be measured in the electromagnetic calorimeters, providing a clean signature for a $W$ decay. It is crucial to distinguish between $W^+$ and $W^-$ since this carries the information on the flavor of the colliding quarks. This is achieved by identifying the charge sign of the high momentum lepton from the $W$ decay, requiring  high resolution tracking in the acceptance of the calorimeters. While the mid-rapidity region is covered by the TPC, the EEMC from 1 to 2 in $\eta$ is currently not covered by trackers in STAR. In order to reliably determine the curvature of the outgoing lepton, and with that determine its charge sign, a device with a spatial resolution of $\sim$ 80 $\mu$m is needed in the forward region. 
The FGT is designed to satisfy these requirements. It is based on GEM technology \cite{Sauli:1997qp}, using a triple GEM configuration similar to the one successfully applied by the COMPASS experiment \cite{Altunbas:2002ds}. The forward tracker will consist of 6 triple GEM disks along the beam pipe with an outer radius of $\sim$ 40 cm and an inner radius of $\sim$ 7 cm.  This ensures that the acceptance of the EEMC is covered over the full extend of the interaction diamond. Each disk will be constructed from 4 quarter sections, requiring GEM foils shaped as 90$^\circ$ wedges with a radius of 40 cm. The readout plane of the detectors will be a two dimensional strip readout with strips in radial and azimuthal direction, with a pitch around 400 $\mu$m. The front-end electronics for this detector will be based on the APV25-S1 chip \cite{French:2001xb}, which will also be used for the IST, significantly reducing development costs for the readout and data acquisition system. For this large-scale project the commercial availability of GEM foils is desirable. A collaboration with Tech-Etch Inc. of Plymouth, MA, USA has been established to develop the production process for these foils. Extensive tests of Tech-Etch produced GEM foils have been performed, both with an X-ray source \cite{Simon:2007sk} and in particle beams at the MTest facility at Fermilab. With a two dimensional readout board with 635 $\mu$m pitch, which is $\sim$ 50\% larger than anticipated for the final design, a spatial resolution of better than 70 $\mu$m was demonstrated, showing that the triple GEM technology satisfies the requirements for forward tracking in STAR.

\section{Summary}

The STAR experiment at RHIC is preparing a comprehensive upgrade of its tracking system over the next few years. These upgrades will enable STAR to achieve topological reconstruction of open charm and open bottom in polarized $p+p$ collisions and in heavy ion collision as well as charge sign identification for electrons and positrons from $W$ decays in $p+p$ collisions at forward rapidities. The upgrades are based on silicon pixel and strip technology and large area triple GEM trackers.

\section*{References}

\bibliographystyle{iopart-num}   
\bibliography{FSimonSTARTracking}

\end{document}